\documentclass[prl,aps,twocolumn,showpacs]{revtex4}

\usepackage{amsmath,graphicx,epsfig}
\begin{document}

\def\pd#1#2{\frac{\partial #1}{\partial #2}}

\title{\bf Solitary wave complexes in two-component mixture condensates}
\author {Natalia G. Berloff}
\affiliation {Department of Applied Mathematics and Theoretical Physics,
University of Cambridge,  Cambridge, CB3 0WA\\
}
\date{4 December 2004}

\begin {abstract} Axisymmetric three-dimensional solitary waves in uniform
  two-component mixture Bose-Einstein condensates are obtained as
  solutions of the coupled Gross-Pitaevskii equations with equal
  intracomponent but varying intercomponent interaction strengths.
   Several families of solitary wave complexes are found: (1) vortex
  rings of various radii in each of the components, (2) a vortex ring in
  one component coupled to a rarefaction solitary wave of the other
  component, (3) two coupled rarefaction waves, (4) either a vortex ring
  or a rarefaction pulse coupled to a localised disturbance of a very
  low momentum. The continuous families of such waves are shown in the
  momentum-energy plane for various values of the interaction
  strengths and the relative differences between the chemical potentials of
  two components. Solitary wave formation, their stability and solitary wave
  complexes in two-dimensions are discussed.
\end{abstract}
\pacs{ 03.75.Lm, 05.45.-a,  67.40.Vs, 67.57.De }
\maketitle
Solitons and solitary waves  represent the essence of many
nonlinear dynamical processes from motions in fluids to energy transfer
along biomolecules, as they define
possible states that can be excited in the system. These are  localised disturbances of the uniform field that are
form-preserving and move with a constant velocity. 
The nonlinear
Schr\"odinger (NLS) equation $i\psi_t+\nabla^2 \psi + \gamma |\psi|^2
\psi = 0$ is canonical and universal equation which is of major
importance in continuum mechanics, plasma physics, nonlinear optics
and condensed matter (where it describes the behaviour of a  weakly
interacting Bose gas and known as the Gross-Pitaevskii (GP)
equation). The reason for its importance and ubiquity is that it
describes the evolution of the envelope $\psi$ of an almost
monochromatic wave in a conservative system of weakly nonlinear
dispersive waves. Similarly,  systems of the coupled NLS equations have
been used to describe  motions and interactions of more than one
wave envelopes in cases when more than one order parameter is needed
to specify the system.  The coupled NLS equations have been
receiving a lot of attention  with  recent
experimental  advances in multi-component
Bose-Einstein condensates (BECs). BECs can excite  various exotic topological
defects and provide a perfect testing ground to investigate their
physics, because almost all parameters of the system can be controlled
experimentally.
Topological defects in two-component BECs 
have  been predicted  theoretically, but there is still no
understanding of what  the complete families of these defects and
solitary waves are,
nor of 
their properties, formation mechanisms and dynamics. It has also been
suggested \cite{volovik}  that 
multi-component BECs  offer the simplest tractable microscopic
models in the proper universality class  of cosmological systems and
solitary waves in multi-component BECs may have their analogs among
cosmic strings. The goal of this Letter is to find and characterise
the families of
solitary waves that exist in  systems of the coupled NLS equation in
three dimensions. The implications of these solitary waves are
wide-ranging, but they will be discussed in the context of
 two-component BECs.

\begin{figure}[t!]
\caption{(colour online) The  dispersion curves of four families of the axisymmetric
  solitary wave solutions of (\ref{ugp}) (A) (red line) $\alpha=0.7$
  and $\Lambda^2=0.1$ ($c_{-}=0.2738$); (B) (black line) $\alpha=0.1$
  and $\Lambda^2=0.1$ ($c_{-}=0.6169$); (C) (green line)  $\alpha=0.5$
  and $\Lambda^2=0.25$ ($c_{-}=0.3317$); (D) (blue line) $\alpha=0.5$
  and $\Lambda^2=0.1$ ($c_{-}=0.3905$).
   The numbers next to the dots give the velocity of
  the corresponding solitary wave. For (D)  (blue line) these are $0.3,
  0.32, 0.34, 0.36, 0.38$. All these solutions are VR-VR complexes
  except for $U=0.58$ on (B) branch which is VR-RP. The top inset
  shows the density isosurface at
  $|\psi_1|^2=\frac{1}{10}\psi_{1\infty}^2$ and
  $|\psi_2|^2=\frac{1}{10}\psi_{2\infty}^2$ for a half of the VR-VR
  complex for $\alpha=0.5$, $\Lambda^2=0.25$ that is moving with
  $U=0.3$. The radii are $b_1=5.194$ and $b_2=4.796$. The density contour plots of this solution are shown in the
  bottom inset.}
\centering
\bigskip
\epsfig{figure=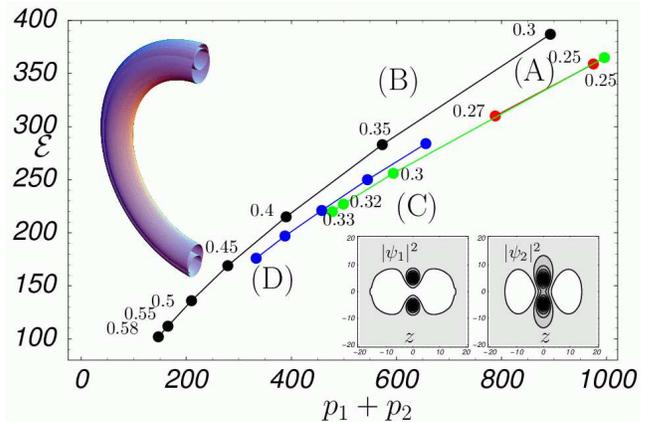, height = 2.2 in}
\label{fig0}
\end{figure}

The simplest example of a multi-component system is a mixture of two
different species of bosons, for instance, ${}^{41}$K-${}^{87}$Rb \cite{mixed}. Since alkali atoms have spin, it is
also possible to make mixtures of the same isotope, but in different
internal spin states, for instance, for ${}^{87}$Rb  \cite{jila}. The
multi-component BECs are far from being a trivial extension of a
one-component BEC and present novel and fundamentally different
scenarios for their excitations and ground state \cite{ex}.
The theory for a mixture of two different bosonic atoms can be
developed similar to that for a one-component condensate whose
equilibrium and dynamical properties  can be accurately 
described by the GP equation \cite{gp} for the wave
function $\psi$ of the condensate
\begin{equation}
i \hbar \frac{\partial \psi({\bf r},t)}{\partial t} = -\frac{\hbar^2}{2 m}\nabla^2\psi({\bf r},t) +  V_0|\psi({\bf r},t)|^2 \psi({\bf r},t),
\label{gp}
\end{equation}
where $m$ is the mass of the atom, $V_0=4\pi\hbar^2a/m$ is the effective interaction between
two particles, and $a$ is the scattering
length.  The GP model has been remarkably
successful in predicting the condensate shape in an external
potential, the dynamics of the expanding condensate cloud and  the motion of
quantised vortices. The family of the solitary waves for (\ref{gp})
was numerically obtained in \cite{jr}. In a momentum-energy ($p{\cal E}$) plot,
the sequence of solitary waves  
has two branches meeting at a cusp where $p$ and ${\cal E}$ simultaneously
assume their minimum values. For each $p$ in excess of the minimum $p_c$, two values
of ${\cal E}$ are possible, and ${\cal E} \to \infty$ as $p \to \infty$ on each branch. 
On the lower (energy) branch the solutions are asymptotic to 
large circular vortex rings. As $p$ and ${\cal E}$ decrease from infinity 
on this branch, the solutions begin to lose their similarity to
vortex rings. Eventually, for a momentum $p_0$ slightly in excess 
of $p_c$, they lose their vorticity, and
thereafter the solutions may better be described as
``rarefaction waves''. The upper branch solutions consist 
entirely of these waves and, as $p \to \infty$, they asymptotically
approach the rational soliton solution of the Kadomtsev-Petviashvili
Type I equation.

For two components,
described by the wave functions $\psi_1$ and $\psi_2$, with $N_1$ and
$N_2$ particles respectively,
 the GP equations
become
\begin{eqnarray}
i \hbar \frac{\partial \psi_1}{\partial t} &=&\biggl[ -\frac{\hbar^2}{2
    m_1} \nabla^2 + V_{11}|\psi_1|^2 +V_{12}|\psi_2|^2\biggr]\psi_1,\nonumber\\
i \hbar \frac{\partial \psi_2}{\partial t} &=&\biggl[ -\frac{\hbar^2}{2
    m_2} \nabla^2 + V_{12}|\psi_1|^2 +V_{22}|\psi_2|^2\biggr]\psi_2, 
\label{two}
\end{eqnarray}
where $m_i$ is the mass of the atom  of the  $i$th condensate,  and the coupling constants $V_{ij}$ are proportional to scattering
lengths $a_{ij}$ via $V_{ij}=2\pi\hbar^2a_{ij}/m_{ij}$, where
$m_{ij}=m_im_j/(m_i+m_j)$ is the reduced mass.

Two-component one-dimensional  BECs have recently
been considered and  various structures have been identified \cite{dd}
such as 
bound dark-dark,  dark-bright, dark-antidark,
dark-grey, etc. complexes.
In higher dimensions, 
  domain walls  \cite{dw} and
skyrmions (vortons) \cite{skyrmions} have been identified by numerical
simulations. 
Numerical simulations of  two-dimensional rotating two-component condensates
were performed \cite{tsubota} and the structure of vortex states
were investigated. A phase diagram in the intercomponent-coupling
versus rotation-frequency plane revealed rich equilibrium structures
such as triangular, square and double-core lattices and vortex
sheets. These simulations give a taste of a rich variety of static
and dynamic phenomena in multi-component condensates.  One would expect the existence of various other families of
solutions many of which have  not yet been  detected. 

In what follows I determine the families of three-dimensional axisymmetric
solitary wave solutions that move with a constant velocity $U$ in {\it uniform}
two-component  mixture BECs. The trap geometry, relevant to experiments,
introduces a harmonic-oscillator potential in (\ref{two}) together
with additional
parameters, places restrictions on studies of solitary waves and their
stability and is irrelevant  in the view of our interest to effects
that occur in large systems. Also,  I believe that additional
physical mechanisms  should be introduced only {\it after} simpler
models are well understood.  Nevertheless, the results I obtain will
be relevant to experiments with a sufficiently shallow trap, so
the linear dimensions of the trap are much larger than the healing length. To reduce the number of parameters in the
system, I will assume that the
intracomponent scattering lengths
and masses
of individual components in the mixture are equal, so that $m_i=m$ and
$a_{ii}=a$, but the intercomponent scattering lengths differ from
$a$. 

To find axisymmetric
  solitary wave solutions moving with velocity $U$ in positive
  $z-$direction, I  solve
\begin{eqnarray}
2iU\frac{\partial \psi_1}{\partial z} &=& \nabla^2 \psi_1 +
(1-|\psi_1|^2-\alpha|\psi_2|^2)\psi_1\nonumber \\
2iU\frac{\partial \psi_2}{\partial z} &=& \nabla^2 \psi_2 +
(1-\alpha|\psi_1|^2-|\psi_2|^2-\Lambda^2)\psi_2,\label{ugp}\\
&&\psi_1 \rightarrow \psi_{1\infty}, \quad \psi_2 \rightarrow \phi_{2\infty}, \quad {\rm as}
\quad |{\bf x}| \rightarrow \infty,\nonumber
\end{eqnarray}
where a dimensionless form of (\ref{two}) is used such that the
distances are measured in units of the correlation (healing) length
$\xi=\hbar/\sqrt{2m\mu_1}$,  the frequencies are measured in units $2\mu_1/\hbar$
and the absolute values of the fields $|\psi_1|^2$ and $|\psi_2|^2$ are
measured in units of particle density $n=\mu_1/V_{11}$. Also present in
(\ref{ugp})  are the parameter of the intercoupling strength
$\alpha =V_{12}/V_{11}$ and the measure of asymmetry  between
chemical potentials $\Lambda^2=(\mu_1-\mu_2)/\mu_1$ (where we assume
that $\mu_1 > \mu_2$). 

The dispersion relation between the frequency $\omega$ and the wave
number $k$ of the linear perturbations ($\propto \exp[i{\bf k} \cdot
{\bf x} - i \omega t]$) around homogeneous states,
$\psi_{i\infty}$, is obtained as
\begin{equation}
\bigl(4\omega^2-k^2(k^2+2\psi_{1\infty}^2)\bigr)\bigl(4\omega^2
-k^2(k^2+\psi_{2\infty}^2)\bigr)
=Pk^4,
\label{omega}
\end{equation}
where $P=\alpha^2\psi_{1\infty}^2\psi_{2\infty}^2$. The condition of dynamic stability
$\omega(k)>0$ gives $\alpha^2<1$
If $\alpha<-1$, the gas is unstable to formation of a
denser state containing both components, while if $\alpha>1$, then the
two components will separate. In what follows I will  consider
$0<\alpha<1$.   The values of the wave-functions of the solitary waves
at infinity in (\ref{ugp})
are given by $\psi_{2\infty}^2=(1-\alpha-\Lambda^2)/(1-\alpha^2)$ and
$\psi_{1\infty}^2=1-\alpha \psi_{2\infty}^2$. In the long-wave limit
($k\rightarrow 0$),
(\ref{omega}) gives two acoustic branches $\omega_\pm \approx c_\pm
k$ with the corresponding sound velocities
$c_\pm=\frac{1}{2}(\psi_{1\infty}^2+\psi_{2\infty}^2\pm\sqrt{(\psi_{1\infty}^2-\psi_{2\infty}^2)^2+4\alpha^2\psi_{1\infty}^2\psi_{2\infty}^2})^{1/2}$.
The solitary waves I seek below are all subsonic, i.e. $U<c_{-}$. This
gives a restriction on the asymmetry parameter $\Lambda^2$: $c_{-}$ is
real only
if $\Lambda^2<1-\alpha$.

\begin{figure}[t]
\caption{(colour online) The  dispersion curves of three families of the axisymmetric
  solitary wave solutions of (\ref{ugp}) with $\alpha=0.1$ and
  $\Lambda^2=0.1$. The numbers next to the dots give the velocity of
  the solitary wave solution. The top (black) branch corresponds to
  VR-VR (VR-RP for $U=0.58$) complexes. The middle (green) branch
  shows $p$ vs ${\cal E}$ for VR-SW complexes and the bottom (red) branch is the
  dispersion curve of SW-VR (SW-RP for $U=0.58$) complexes. The radii of the vortex
ring solutions are shown on the upper inset as a function of $U$: the
  two top (red and green) lines give $b_1$ and $b_2$ correspondingly
  for VR-VR complexes. The two bottom (black and blue) lines represent
  $b_1$ in VR-SW complex and $b_2$ in SW-VR complex
  correspondingly. The bottom inset shows 3D plots of $|\psi_i(s,z)|^2$
  of the SW-RP complex moving with the velocity $U=0.58$.
}
\centering
\bigskip
\epsfig{figure=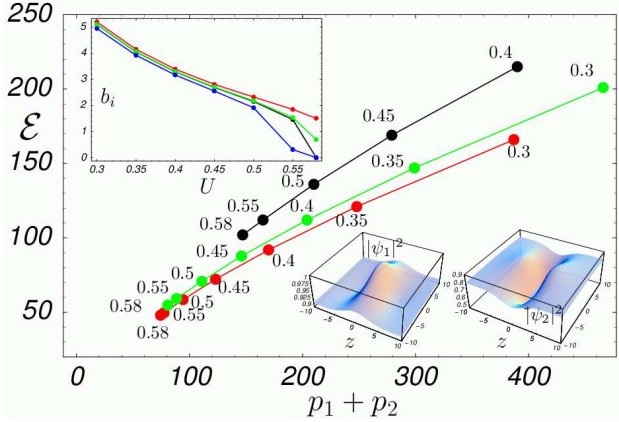, height = 2.2 in}
\label{fig1}
\end{figure}

Each solitary wave complex that belongs to a family of the solitary wave solutions for a
chosen set of $(\alpha, \Lambda^2)$ will be characterised by its
velocity, $U$, vortex
radii $b_i$, momenta ${\bf p}_i=(0,0,p_i)$,
and energy ${\cal E}$. The momentum (or impulse) of the $i-$th
component is ${\bf p}_i=\frac{1}{2
  i}\int [(\psi_i^*-\psi_{i\infty})\nabla\psi_i
  -(\psi_i-\psi_{i\infty})\nabla\psi_i^*]\,dV$. The reasons for
replacing a more customary
  defined momentum $\hat {\bf p}_i=\frac{1}{2
  i}\int \psi_i^*\nabla\psi_i
  -\psi_i\nabla\psi_i^*\,dV$  with the convergent integrals ${\bf p}_i$ were
  spelled out in \cite{jr} for a one component
  GP equation. Also, similar to \cite{jr}, we form the energy, ${\cal
    E}$, by subtracting the energy of an undisturbed system of the same
  mass for which $\psi_i=const$  everywhere, from the energy of the
  system with a solitary wave, so that the energy of the system becomes
\begin{eqnarray}
{\cal E}&=&\frac{1}{2}\int|\nabla \psi_1|^2+|\nabla \psi_2|^2\,
  dV \nonumber \\
&+&\frac{1}{4}\int(\psi_{1\infty}^2-|\psi_1|^2)^2
+(\psi_{2\infty}^2-|\psi_2|^2)^2\, dV\label{e1} \\
&+&\frac{\alpha}{2}\int(\psi_{1\infty}^2-|\psi_1|^2)(\psi_{2\infty}^2-|\psi_2|^2)\,
  dV.\nonumber
\end{eqnarray}
By performing the variation $\psi_i\rightarrow \psi_i+\delta\psi_i$ in
the integrals for ${\cal E}$ and $p_i$, discarding surface integrals
that vanish provided $\delta\psi_i\rightarrow 0$ for $|{\bf
  x}|\rightarrow \infty$ and making use of (\ref{ugp}), we obtain
$\delta {\cal E}=U\delta(p_1+p_2)$, or $U=\partial {\cal E}/\partial (p_1+p_2)$,
  where the derivative is taken along the solitary wave sequence. The same
  expression is obeyed by the sequences of classical vortex rings in
  an incompressible fluid and by the solitary waves of \cite{jr}.
By applying some algebraic manipulations that involve some integration by
parts, the following integral properties can be  established:
\begin{eqnarray}
&&{\cal
  E}=\int\left|\pd{\psi_1}{z}\right|^2+\left|\pd{\psi_2}{z}\right|^2\,
dV, \label{e3}\\
&&{\cal
  E}=\frac{1}{2}\int(1-|\psi_1|^2-\alpha|\psi_2|^2)(2\psi_{1\infty}^2\!\!-\!\psi_{1\infty}(\psi_1+\psi_1^*))\,dV \nonumber\\
&&+ \frac{1}{2}\int(1-\alpha|\psi_1|^2-|\psi_2|^2)(2\psi_{2\infty}^2\!\!-\!\psi_{2\infty}(\psi_2+\psi_2^*))\,dV
\label{e4}\\
&&U(p_1+p_2)=\frac{1}{4}\int(1-|\psi_1|^2-\alpha|\psi_2|^2)\nonumber
\\
&&\hskip 0.6 in\times(3\psi_{1\infty}^2-\psi_{1\infty}(\psi_1+\psi_1^*)-|\psi_1|^2)\,dV \nonumber\\
&&\hskip 0.3 in+\frac{1}{4}\int(1-\alpha|\psi_1|^2-|\psi_2|^2)\nonumber \\
&&\hskip 0.6 in\times(3\psi_{2\infty}^2-\psi_{2\infty}(\psi_2+\psi_2^*)-|\psi_2|^2)\,dV.
\label{p3}
\end{eqnarray}
These were used as checks on numerical accuracy of the solutions I
obtain next.
 
The axisymmetric solitary waves are found by re-writing (\ref{ugp}) in cylindrical
coordinates $(s,\theta,z)$ for the deviations from the 
solutions at infinity $\Psi_i=\psi_i-\psi_{i\infty}$. 
Stretched variables $z'=z$ and $s'=s\sqrt{1-2 U^2}$ were introduced and
 the infinite domain was mapped onto the box
$(0,\frac{\pi}{2})\times(-\frac{\pi}{2},\frac{\pi}{2})$ using the transformation
$
\widehat z=\tan^{-1}(D z')
$ and $
\widehat s=\tan^{-1}(D s'),$
 where  $D\sim 0.4-0.5$.
Transformed equations (\ref{ugp}) were expressed in second-order finite
difference form using $100^2$ grid points, and the resulting nonlinear equations were solved by
Newton-Raphson iteration procedure using banded matrix 
 linear solver based on bi-conjugate gradient stabilised iterative method with
preconditioning. 
As
$\alpha \rightarrow 0$, two components become uncoupled, so the
solitary wave sequence for each component is following the dispersion
curve of the one-component GP equation (\ref{gp}) with vortex rings (VRs)
becoming rarefaction pulses (RPs) as $U$ increases, with energy
and momentum being appropriately scaled by the densities at
infinity. For $\alpha \ne 0$, different components become RP at
different critical values of $U$, so variety of complexes become 
possible. Table 1 gives  an example of various transitions from one complex
to another in the system with $\alpha=0.05$ and $\Lambda^2=0.1$. Notice that the
radii of the vortex rings in VR-VR complexes differ with $b_2
\rightarrow b_1$ as $U \rightarrow 0$. Also notice, that there is 
a cusp in the dispersion curve ${\cal E}$ vs $p_1+p_2$, since the solitary
wave complex moving with $U=0.63$ belongs to the upper branch.

\bigskip 

\noindent {\bf Table 1.} {\footnotesize The
velocity, $U$,
  energy, ${\cal E}$, momenta, $p_i$ and radii, $b_i$, of the
  solitary wave solutions of (\ref{ugp}) with $\alpha=0.05$ and
  $\Lambda^2=0.1$. The sequence terminates at $U = c_{-}\approx
  0.646.$}
\medskip

\begin{tabular}{c  c c c c c c}
\hline
$U$ & ${\cal E}$  & $p_1$ & $p_2$ & $b_1$ & $b_2$ & complex \\
\hline
0.55 & 118 & 92.3 & 80.8 & 1.83 & 1.50 & VR-VR \\
0.58 & 107 & 81.0 & 72.6 & 1.47 & 0.57 & VR-VR \\
0.60 & 101 & 74.9 & 69.6 & 1.12 & --   & VR-RP \\
0.63 & 102 & 66.2 & 79.0 &  --  & --   & RP-RP \\
\hline
\end{tabular}
\medskip

As the interaction strength $\alpha$  increases for fixed  $\Lambda^2$ and
$U$, the
radii in VR-VR complexes don't change much, although energy and
momentum decrease significantly. We can compare two such
solutions for $\Lambda^2=0.1$ and $U=0.3$: the VR-VR complex for
$\alpha=0.1$ has ${\cal E}=387$, $p_1=484$, $p_2=410$, $b_1=5.197$
and $b_2=5.096$, whereas for $\alpha=0.5$ these values are ${\cal E}=284$, $p_1=385$, $p_2=271$, $b_1=5.196$
and $b_2=5.093$. On the other hand, if $\alpha$ and $U$ are kept
constant and the asymmetry parameter $\Lambda^2$ increases, the radius of the
vortex ring in
the
second component 
 decreases  (if
$\Lambda^2=0.5$, $\alpha=0.1$, $U=0.3$, then ${\cal E}=284$, $p_1=503$, $p_2=160$, $b_1=5.182$
and $b_2=4.363$). Fig. \ref{fig0} shows the dispersion curves of 
families of the axisymmetric solitary wave complexes for a variety of
$(\alpha, \Lambda^2)$.

In addition to the VR-VR, VR-RP, and RP-RP complexes that I have just
described, two other families of the solitary waves came as a
surprise. In contrast with the VR-VR, VR-RP, and RP-RP complexes,
in which the solitary wave in each component possesses a large
momentum, the total momentum of the system in the new families  is almost
entirely belongs to one component with the disturbance of the other
component moving almost entirely through the coupling with the mobile
component. I will call this disturbance a ``slave wave'' (SW). This
gives rise to two other complexes such as VR-SW and RP-SW. Notice that
the role of the components can be reversed giving SW-VR and SW-RP
complexes as well. To indicate how the characteristics of these
complexes change with increasing intercomponent interaction strength
$\alpha$, the parameters of these solutions are given in Table 2 for
$U=0.3$ and $\Lambda^2=0.1$. Fig.\ref{fig1} shows the dispersion
curves  of all three
families of the axisymmetric solitary wave solutions for $\alpha=0.1$
and $\Lambda^2=0.1$.

\smallskip

\noindent {\bf Table 2.} {\footnotesize The intercomponent interaction strength, $\alpha$,
  energy, ${\cal E}$, momenta, $p_i$ and radii, $b_i$, of the
  solitary wave solutions of (\ref{ugp}) with $U=0.3$ and
  $\Lambda^2=0.1$. The sequence terminates ($c_{-}=U$) at $\alpha\approx
  0.65.$}
\medskip

\begin{tabular}{c  c c c c c c}
\hline
$\alpha$ & ${\cal E}$  & $p_1$ & $p_2$ & $b_1$ & $b_2$ & complex \\
\hline
0.2 & 178 & 414 & 1.52 & 4.977 & -- & VR-SW \\
0.4 & 142 & 330 & 7.76 & 4.673 & -- & VR-SW \\
0.6 & 119 & 268 & 35.1 & 4.122 & --   & VR-SW \\
\hline
\end{tabular}

\begin{tabular}{c  c c c c c c}
\hline
$\alpha$ & ${\cal E}$  & $p_1$ & $p_2$ & $b_1$ & $b_2$ & complex \\
\hline
0.2 & 140 & 1.12 & 330 & -- & 4.787 & SW-VR \\
0.4 & 96.7 & 4.89 & 237 & -- & 4.354 & SW-VR \\
0.6 & 63.0 & 17.5 & 174 & -- & 2.986   & SW-VR \\
\hline
\end{tabular}
\medskip

Finally, when the solitary waves are found, there is a need to
elucidate the following topics. {\it Stability.}  The Derrick-type
argument used in \cite{jr} can be applied together with the integral
identities (\ref{e1})-(\ref{p3}), to indicate the stability of 
solitary waves as long as $\partial U/\partial (p_1+p_2)
<0$ along the family. This condition is satisfied for all solitary waves discussed here
except for the last entry of Table 1. {\it Solitary waves in
  2D}. Similarly to 3D complexes, solitary waves in 2D consist of
the combinations of a  pair of two vortices with opposite circulation
(VP), 2D rarefaction pulse, or a ``slave wave,'' giving rise to
VP(RP)-VP(RP) and  SW-VP(RP) complexes. {\it Formation.} The
mechanisms of solitary wave nucleation in two-component condensates
are similar to those in one-component condensates \cite{ngb}. In
particular, the moving objects (ions, laser beams etc) shed VR(RP)-VR(RP) complexes in 3D and
VP(RP)-VP(RP) complexes in 2D, when the local speed of sound is
reached somewhere on the surface of the object.
I will ellaborate on these issues in details elsewhere.

The support from  NSF grant DMS-0104288 is 
acknowledged. 

\end{document}